# Ab-initio study of the stability and electronic properties of wurtzite and zinc-blende BeS nanowires

Somayeh Faraji[1] and Ali Mokhtari[1,2*]

[1]Simulation Laboratory, Department of Physics, Faculty of Science, Shahrekord University, Shahrekord, Iran

[2]Nanotechnology Research Center, Shahrekord University, Shahrekord, Iran

**Abstract**

In this work we study the structural stability and electronic properties of the Beryllium sulphide nanowires (NWs) in both zinc blende (ZB) and wurtzite (WZ) phases with triangle and hexagonal cross section, using first principle calculations within plane-wave pseudopotential method. A phenomenological model is used to explain the role of dangling bonds in the stability of the NWs. In contrast to the bulk phase, ZB-NWs with diameter less than 133.3Å are found to be less favorable over WZ-NWs, in which the surface dangling bonds (DBs) on the NW facets play an important role to stabilize the NWs. Furthermore, both ZB and WZ NWs are predicted to be semiconductor and the values of the band gaps are dependent on the surface DBs as well as the size and shape of NWs. Finally, we performed atom projected density-of states (PDOSs) analysis by calculating the localized density of states on the surface atoms, as well as on the core and edge atoms.

## 1. Introduction

In the past decades, nanoscience and nanotechnology has been making significant progress, and its effects on every field have been acknowledged in the world [1]. The nanomaterials have attracted enormous attention due to their size-dependent unique mechanical, physical and chemical properties. Among them, one-dimensional semiconductor nanostructures (such as nanowires, nanorods, and nanotubes) are promising candidates for technological nanoscale electronic, optoelectronic and photonic applications [2-4]. In these nanostructures, electrons are confined in two directions with a small cross-section area and a large surface-to-volume ratio; therefore they present interesting electronic and optical properties due to the quantum confinement effects.

Wide band gap semiconductor compounds have a great potential for applications with light-emitting and laser diodes (LEDs and LDs) in the visible region of the spectrum [5,6]. These materials are characterized by different degrees of covalent and ionic bonding and thus offer a wide range of physical properties. Among them, beryllium chalcogenide BeS, BeSe and BeTe have attracted increasing attention in the past few years [7-27]. Due to the highly toxic nature of these compounds, only few experimental studies are documented [28-33]. Consequently most of the literatures reported on this topic are theoretical researches. The ZB and WZ phases are the most common crystal structures of these compounds.

---

[*] Author to whom any correspondence should be addressed. E-mail: mokhtari@sci.sku.ac.ir

In the present work we have investigated the structural stability and electronic properties of the pristine ultra-thin BeS nanowiers in both ZB and WZ phases and also the role of the surface dangling bonds (DBs) on nanowire facets. The beryllium sulphide compound has high thermal and low electrical conductivity, high melting point and hardness. This compound is partially ionic semiconductor with large and indirect band gap and is stable in the ZB structure at ambient conditions. This compound is promising candidate material for blue-green laser diodes and laser-emitting diodes [32]. The first experimental study on this matter was performed by Zachariasen [28] who demonstrated that it crystallizes in the ZB structure and measured its lattice constant. To the best of our knowledge, no theoretical and experimental results have yet been reported about the structural stability and physical properties of BeS nanowires. Our calculations with diameter less than 18Å indicate that the ZB nanostructures are less favorable over the WZ nanowires.

Section 2 gives an outline of the computational method and some important parameters. Results and discussion concerning structural stability, electronic properties and role of the dangling bonds, are in section 3. Section 4 summarizes the conclusions.

## 2. Computational details

First-principle calculations have been performed using pseudopotential method, as implemented in the QUANTUM ESPRESSO/ PWSCF package [34], within density functional theory (DFT) [35,36]. The exchange-correlation functional was approximated using the Perdew-Burke-Ernzerhoff [37] form of the GGA. The electron-ion interaction was described by ab-initio ultrasoft pseudopotentials [38]. The total energy has been obtained by solving the standard Kohn-Sham (KS) equations using self-consistent method. The wave-functions (densities) of electrons were expanded on a plane wave basis set up to a kinetic energy cutoff of 40 Ryd (160 Ryd). The Brillouin zone was sampled using up to eight (1×1×8) **k**-points within Monkhorst-Pack scheme [39] along the nanowire axis. The criterion of convergence for the energy was 0.0001 Ry and the maximum force allowed on each atom was 0.002 Ry/a.u. All nanostructures have been treated within a supercell geometry using the periodic boundary conditions. Vacuum spacing was arranged so that the minimum distance between two atoms in adjacent unit cells were about 10.3 Å, provided that atoms have negligible interaction at that far distance.

## 3. Results and discussions

3. 1. Structural stability of nanowires

The lattice constants and internal parameter are first optimized for the bulk BeS in the both ZB and WZ phases. The optimized structural parameters are $a$=3.43 Å, $c$=5.66 Å, $u$=0.374 for the WZ and $a$=4.87 Å for the ZB structure, which are in good agreement with the experimental value and the results of theoretical works [20-24].

The cohesive energy (per atom) is defined as the difference between the average energy of isolated atoms and the crystal energy per atom. In order to obtain an accurate value for the cohesive energy, the energy calculations for isolated atoms and crystal, must be performed at the same level of accuracy. To fulfill such a requirement, the energy of an isolated atom was computed by considering a large cell containing just one atom. The size of this cube was chosen sufficiently large so that the energy convergence with respect to the size of the cube was less than 0.0001Ryd. Our results (3.928 and 3.922 eV respectively for ZB and WZ phases) for the bulk cohesive energy indicate that the ZB structure is more stable, which is also in accordance with the other theoretical work [20].

The ZB nanowires have been constructed along the [111] direction (with hexagonal cross section and ABCABC… arrangements in three double layers) in which the periodicity length is $\sqrt{3}a$, where $a$ is the bulk lattice parameter in ZB phase. These nanostructures are labeled as ZB-H($l, m, n$) where $l$, $m$, and $n$ represent the number of atoms in each double layer [40]. The atomic structure of the WZ nanowires has been founded along the [0001] direction. There are two kinds of ABAB… arrangement for the WZ nanowires with triangle or hexagonal cross section, which are labeled with WZ-T($l, m$) and WZ-H($l, m$) respectively ($l$ and $m$ indicate the number of atoms in each double layer).

For each pristine nanowire, the atomic positions were initially arranged from the bulk structure using the supercell approach and then their optimum values are obtained by relaxing the atomic positions in the force direction using a standard Broyden-Fletcher-Goldfarb-Shanno (BFGS) [41] quasi-Newton method. Throughout relaxation, we used a Gaussian expanding with smearing parameter of 0.01Ryd for Brillouin-zone integrations to eliminate convergence problems due to fractionally occupied surface states. Top views of the relaxed nanostructures are shown in figure 1 for both ZB and WZ phases.

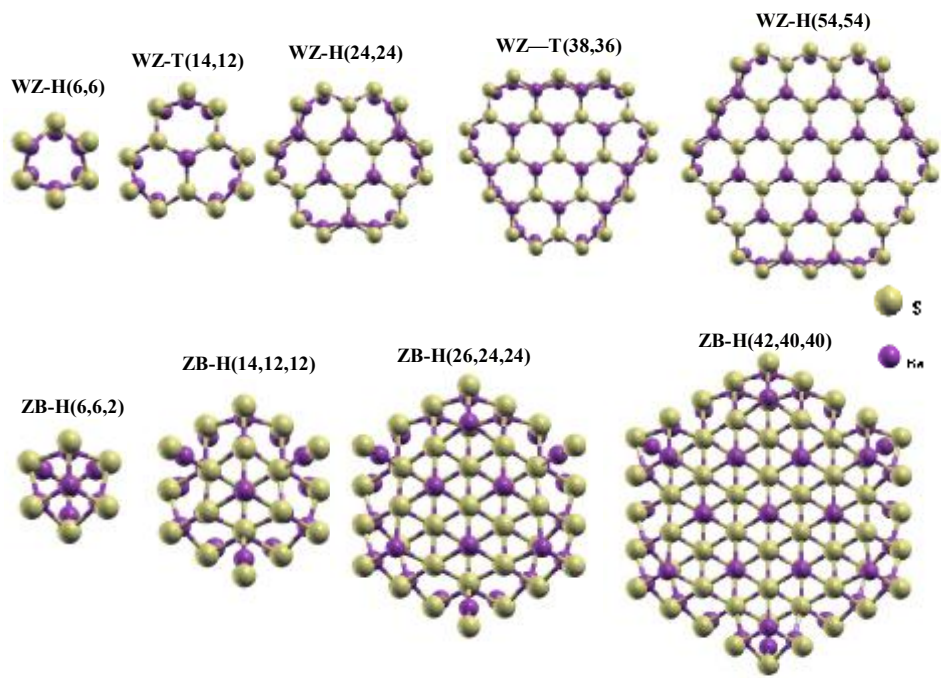

Figure 1: (Colour online) Top view of the relaxed BeS nanowires in WZ-H, WZ-T and ZB-H structures.

The $c$ lattice parameters of NWs were optimized by calculating (self-consistent) the energies at different values of $c$ and fitting the results with a parabolic curve. We have obtained the values of 5.49, 5.61, 5.64 and 5.66Å for WZ and 8.69 and 8.46 Å for ZB NWs at smallest diameters respectively for $c$ lattice constant. Our results indicate that by increasing the diameter, the $c$ parameter approaches quickly to the bulk value. Therefore, we have abandoned the $c$ optimization for the larger diameters and only used the bulk values.

In order to investigate the energetic stability of the BeS nanowires, we can express the cohesive energy (per atom) of nanowires as follows [40]:

$$E_c^{NW}(WZ) = E_c^{bulk}(WZ) - \frac{N_{DB1}}{N_{tot}} E_{DB1}(WZ), \qquad (1)$$

$$E_c^{NW}(ZB) = E_c^{bulk}(ZB) - \frac{N_{DB1}}{N_{tot}} E_{DB1}(ZB) - \frac{N_{DB2}}{N_{tot}} E_{DB2}(ZB). \qquad (2)$$

where $E_c^{bulk}$ is the cohesive energy per atom of the bulk BeS, $E_{DBi}$ and $N_{DBi}$ ($i$=1,2) are the energy and number of DBs respectively and $N_{tot}$ is the total number of atoms in the nanostructures. We have obtained 0.287, 0.353 and 0.556 eV respectively for $E_{DB1}(WZ)$, $E_{DB1}(ZB)$ and $E_{DB2}(ZB)$. Using these data and above formula, we have estimated the cohesive energies of nanostructures for larger diameters up to 180Å by applying the extrapolation method. The behaviour of the calculated and extrapolated cohesive energies respect to the diameter of nanostructures is plotted in figure 2.

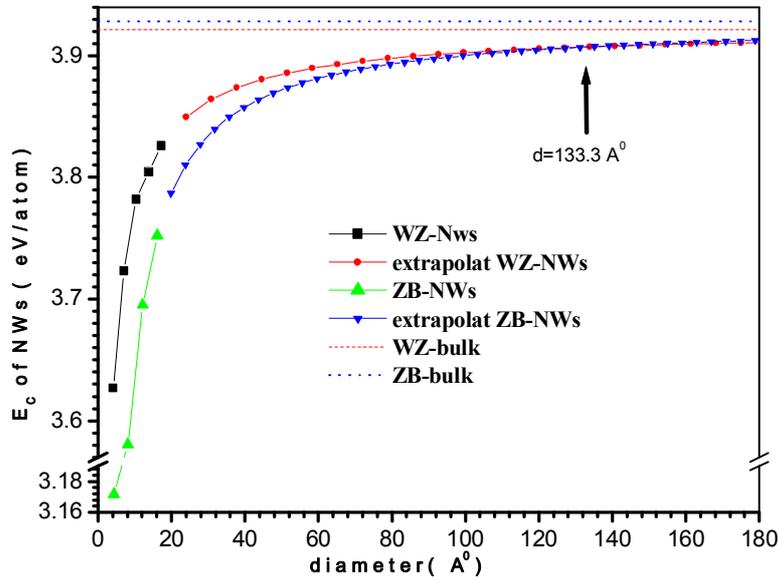

Figure 2: (Colour online) Cohesive energy ($E_c$) of the pristine BeS nanowires as a function of nanowire diameter. The blue and red horizontal line indicates the bulk cohesive energies of ZB and WZ respectively. Green up-triangles and black squares represent our calculated cohesive energy of ZB-NWs and WZ-NWs with diameters less than 18 Å. The extrapolated results obtained from Eqs. (1) and (2) are shown by blue down-triangles and red circles.

The main features to note from the above calculations are as follows:

In the diameters less than 133.3 Å, the WZ-NWs are energetically more favorable than the ZB-NWs.

The DBs ratio is defined as the number of surface DBs generated on the nanostructure facets divided to the total number of atoms in nanostructure. The value of DBs ratio decreases with enhancement the diameter of nanostructures (figure 3).

In contrast to the bulk system, the WZ nanostructures are more stable than the ZB nanostructures. This behaviour is consistent with the trend that smaller DB energy and DBs ratio leads to larger cohesive energy. Similar behaviours were reported for other NWs [40, 42-48]

In the relaxed nanowires, Be atoms move inward and S atoms move outward (figure 1). We also have seen that after relaxation, diameter of nanowires increases and bond length of Be-S on the surfaces decreases.

It is obvious from the top view of the optimized WZ and ZB NWs that most of atomic relaxations take place on the surface and edge atoms. Therefore the large surface-to-volume ratio causes considerable deviation from the bulk properties.

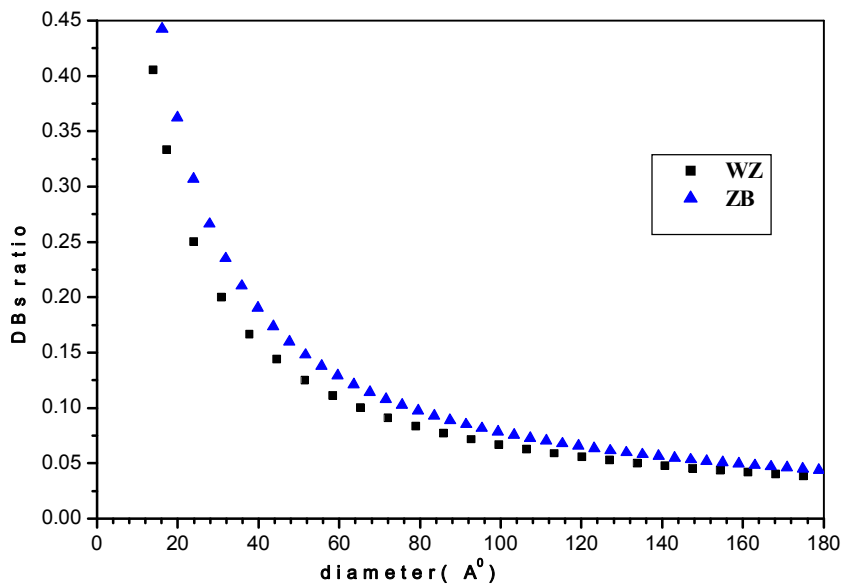

Figure 3: (Colour online) The behaviour of DBs ratio as a function of NW diameter. The blue triangle and black squares indicates the DBs ratio of ZB and WZ nanowires respectively.

3. 2. Electronic properties

At first, the bulk band structures are calculated for both phases. The results show that both phases have indirect band gap. In our theoretical calculations, the values of band gap are predicted about 3.7 and 4 eV respectively for the ZB and WZ bulk, which are smaller than the experimental value (5.5eV for the ZB) and in agreement with other theoretical results [15]. It should be noted that the energy gap is usually underestimated in the DFT by 40-50% than the experimental value [49].

In order to investigate the effects of DBs and size of the NWs on the electronic properties, we have calculated the band structures of NWs using self consistent solving of Kohn-Sham equations by applying the optimized structural parameters. Our results of energy gaps respect to the diameter of NWs ($d_{NW}$) are shown in figure 4 for both phases and compared to the bulk values. In the ZB phase of BeS-NWs the value of the band gap decreased by enhancement of the diameter of NWs. Similar behaviour has been reported for the ZnS nanostructure in the ZB phase [45, 51]. We will explain this unusual trend using the partial density of states. For the WZ-NWs, the surface DB states and the size of NWs determine the band character and energy gap in the small diameters, but the value of the band gap is dependent on only the NW size in larger diameters. The similar behaviour is reported for InP-NWs [40].

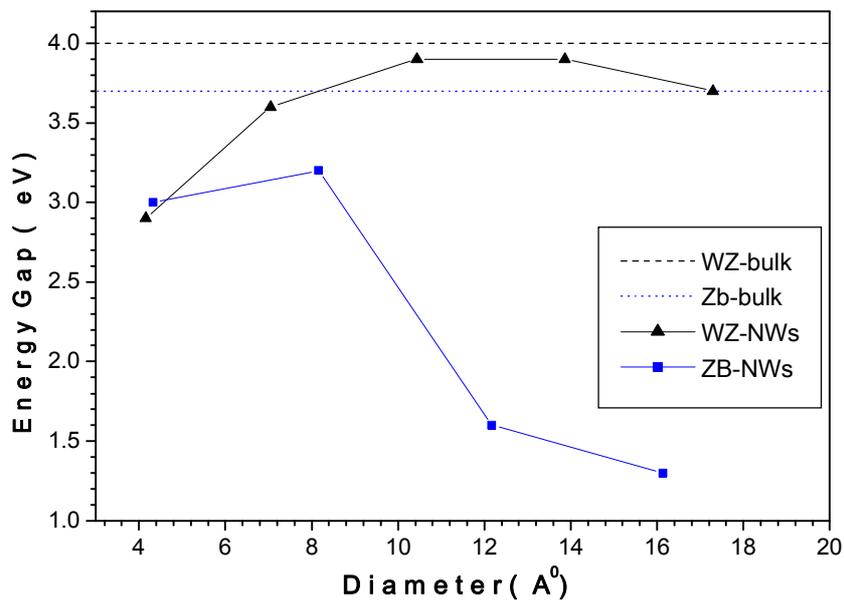

Figure 4: (Colour online) Calculated band gaps for ZB (blue squares) and WZ (black triangles) nanowires as a function of the diameter. The blue dotted and black dashed horizontal lines represent the bulk band gaps of BeS in ZB and WZ phases, respectively.

To analyze the behaviour of the band gaps and to further study the electronic properties of these nanostructures, the total (whole cell contribution) density of states (figure 5) and the contribution of the core, surface and edge atoms are calculated using tetrahedron method [50]. The fundamental points to note from these calculations are as follows:

For the WZ-NWs, the valence bands are separated in two sub-bands that are labeled starting from the top (Zero energy) as VB1 and VB2. The widths of these sub-bands are increased and the internal gaps (the energy gap between VB2 and VB1) between them decreased by increasing the diameter of WZ-NWs. The contributions of the core and surface states in total DOS are calculated for s and p orbitals of S atoms and s orbitals of Be atoms. The results of the WZ-H(54,54) structure are reported as a sample in figure 6. It is evident from this figure that the p orbitals of S atoms hybridize with the s states of Be atoms in the VB1, while the VB1 is formed by overlapping of s orbitals of Be and S atoms. The average energy of the surface states in both VB1 and VB2 parts is larger than the corresponding average energy of core states. This behaviour is consistent with the presence of dangling bonds in the surface of the NWs.

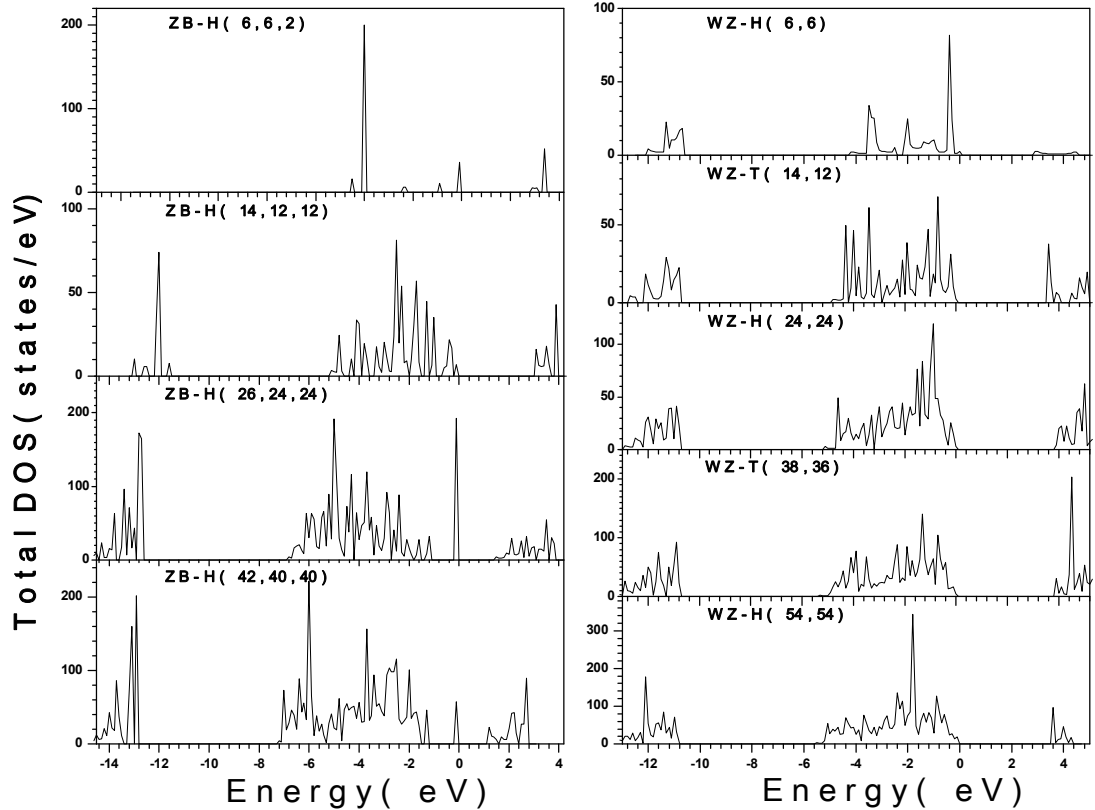

Figure 5: Total electronic densities of states (DOSs) of WZ and ZB NWs. The highest occupied states are adjusted on zero of the energy axis.

There are three kinds of states in the ZB-H nanostructures: core, surface and edge states of Be and S atoms. The contributions of different orbitals in these states are calculated and shown as a sample for the ZB-H(42,40,40) nanostructures in the figure 6. It is evident from this figure that the contributions of the narrow edge states are considerable respect to the surface and core states and also their contributions are dominant in the highest occupied states near the fermi level. The behaviour of the energy gap (figure 4) and partial density of states respect to the diameter of the ZB-NWs is quite different from those of the WZ-NWs. The physical origin of this unusual trend can be treated as four different reasons: the presence of double-dangling bonds in the edge of ZB nanostructures, numerous dangling bonds, quantum confinement effects and high value for the average energy of DBs.

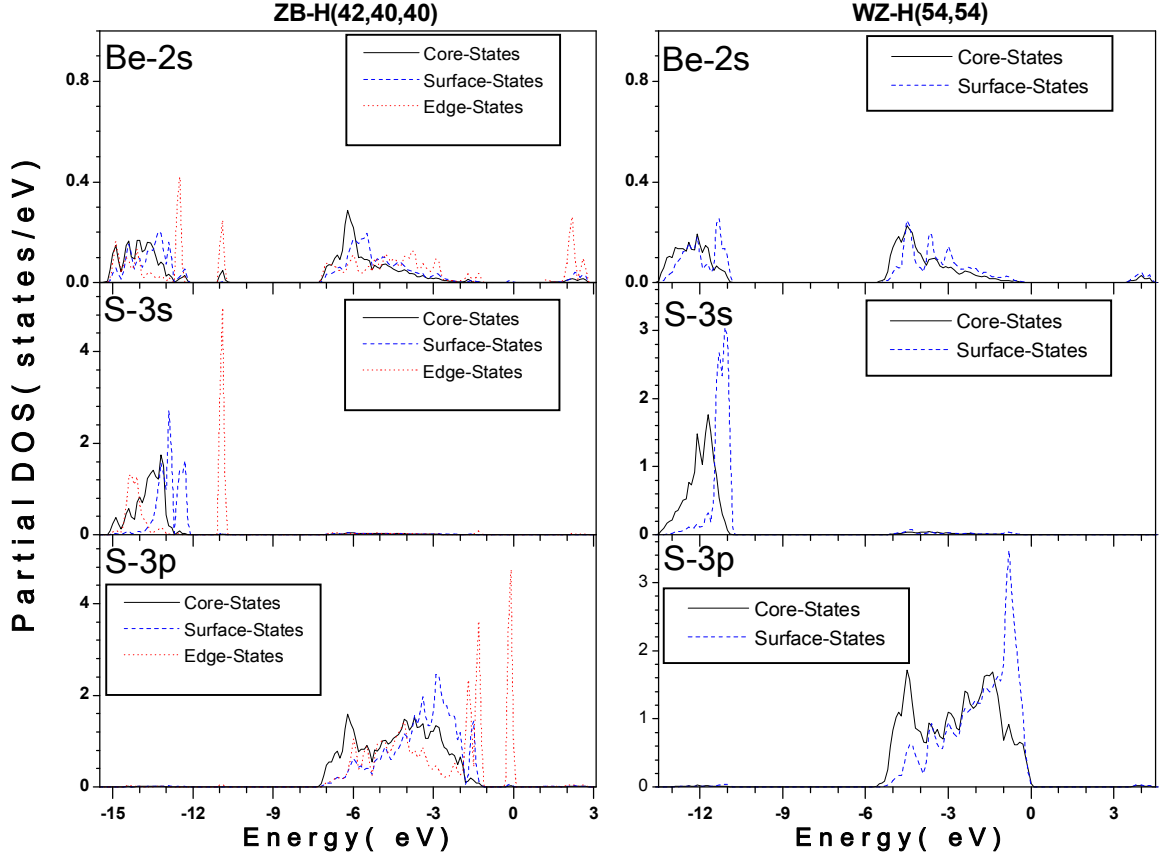

Figure 6: (Colour online) Partial density of states (PDOSs) of the ZB-H(42,40,40) and WZ-H(54,54) NWs. PDOSs for s and p orbitals of S atoms and s orbitals of Be atoms are shown as solid (black) and dashed (blue) lines for core and surface states respectively. The dotted (red) lines indicate the PDOSs of edge states in each mentioned orbital of ZB-H(42,40,40) NWs.

5. Conclusions

We have studied the stability and electronic properties of the pristine BeS NWs, using *ab initio* total energy calculations. We have considered both ZB and WZ phases respectively at [111] and [0001] directions in NW diameters less than 18 Å. We found that WZ-NWs are more stable than ZB-NWs. In order to predict the stability of NWs at larger diameters, we have used a phenomenological model and obtained the DBs energy on the surfaces and edges of nanostructures. By applying the values of DBs energies in the model, we have extrapolated the cohesive energy in larger diameters. We have found out that WZ-NWs with diameter less than 133.3 Å are more stable than the ZB nanostructure. The average DBs energy (0.4545) of the ZB-NWs is larger than the corresponding value (0.287 eV) for other phase. This result is in accordance with the stability of WZ-NWs. In order to explain the behaviour of the structural stability, we have calculated the electronic properties of both phases of these nanostructures. The obtained results indicate that the behaviour of the ZB-NWs is quite different from those of the WZ-NWs.

## 6. Acknowledgment

The authors acknowledge gratefully support of the Shahrekord University for this research. This work was performed in the simulation laboratory of physics department under project number 122-4774.